\documentclass[reprint, amssymb, amsmath, aip, cha]{revtex4-1}
\usepackage[version=3]{mhchem}
\usepackage[T1]{fontenc}
\usepackage{amsmath}
\usepackage{amssymb}
\usepackage{units}
\usepackage{graphicx}
\usepackage{psfrag}
\usepackage{tabularx}
\usepackage{booktabs}
\usepackage{bigstrut}
\usepackage{epstopdf}
\usepackage[caption=false]{subfig}

\begin{document}

\title{Electron Transport in Molecular Junctions with Graphene as Protecting
Layer} 
\date{\today}
\author{Falco H\"user}
\affiliation{Nano-Science Center
  and Department of Chemistry\\
  University of Copenhagen, 2100 K{\o}benhavn {\O}, Denmark}
\affiliation{Current address: Office for Innovation and Sector Services,
  Technical University of Denmark, 2800 Kgs. Lyngby, Denmark}
\author{Gemma C. Solomon}
\email{gsolomon@nano.ku.dk} \affiliation{Nano-Science Center
  and Department of Chemistry\\
  University of Copenhagen, 2100 K{\o}benhavn {\O}, Denmark}


\begin{abstract}
We present \textit{ab-initio} transport calculations for molecular junctions
that include graphene as a protecting layer between a single molecule and gold
electrodes. This vertical setup has recently gained significant interest in
experiment for the design of particularly stable and reproducible devices.
We observe that the signals from the molecule in the electronic transmission
are overlayed by the signatures of the graphene sheet, thus raising the need
for a reinterpretation of the transmission. On the other hand, we see that our
results are stable with respect to various defects in the graphene. For weakly
physiosorbed molecules, no signs of interaction with the graphene are
evident, so the transport properties are determined by offresonant
tunnelling between the gold leads across an extended structure that includes
the molecule itself and the additional graphene layer. Compared with pure gold
electrodes, calculated conductances are about one order of magnitude lower due
to the increased tunnelling distance. Relative differences upon changing the
end group and the length of the molecule on the other hand, are similar.
\end{abstract}


\maketitle

\section{Introduction\label{sec:intro}}
\begin{figure}[t]
\begin{centering}
\includegraphics[width=\columnwidth,clip=]{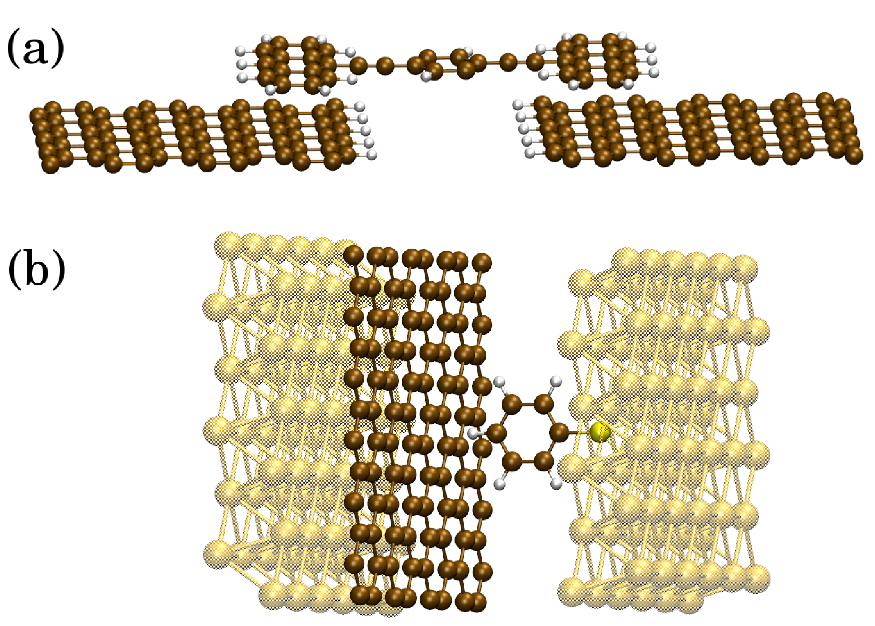}
\par\end{centering}
\caption{Comparison of (a) planar molecular junction with graphene nanoribbons
as electrodes (b) vertical molecular junction with gold electrodes and graphene
as soft top contact layer.}
\label{fig:device}
\end{figure}
Today, molecular electronics is at a crossroads: While charge transport
at the single molecule level has been studied extensively over the past couple
of decades both in theory and experiment, the fabrication of efficient and
longlasting devices at nanoscale to replace convential semiconductor based
electronics seems as out of reach as ever.
\cite{Song11, Kiguchi13, Sun14, Schwarz14, Nichols15}
Traditionally, metal electrodes made of gold have been favoured for the design
of molecular junctions. Though they provide a suitable test bed for
transport measurements and the molecule-metal interface is well understood at
the theoretical level, they do not seem viable for practical applications.
One fundamental issue is that most fabrication techniques lead to
large uncertainties in the junction geometry, resulting in broad variations in
measured conductances. This also means that two different junctions are not
necessarily comparable (and even less with theoretical models).
Furthermore, evaporation of metals onto self-assembled monolayers (SAMs) has
turned out to be invasive and can cause short circuits, thus destroying the
junction.\cite{deBoer04, Lau04} In fact, the device yield with this fabrication
technique is very low.
Finally, limitations in design in the top-down approaches make new
functionalitites difficult to incorporate.

Recently, graphene has been proposed as an alternative soft top contact between
SAMs and metal electrodes.\cite{Wang11, Li12, Petersen12, Parui15}
This material is chemically stable and posesses outstanding electronic and
mechanical properties, making it a perfect candidate for durable and
reproducible devices. It protects the molecular layer from reorganization
and penetration by metal atoms, giving high yield, good operational
stability and long device lifetimes. Since graphene is transparent in the
range of visible light, it can also be used for optoelectronic molecular
switches.\cite{Seo13, Li13} In addition, the incorporation of a back gate
electrode seems achievable.
Most importantly, however, from a fundamental point of view is that due to
its robustness and flexibility, graphene provides an almost perfect interface.

Previous research on molecular junctions including graphene has focused on a
lateral device layout using nanoribbons as leads.
\cite{Zheng10, Carrascal12, Fainberg13}
In these junctions, molecules can form covalent C-C bonds with the
electrodes allowing for direct injection into the molecular backbone
\cite{Nikolic12, Cao13} or interact via $\pi$-$\pi$ stacking of aromatic rings,
\cite{Prins11, Ullmann15}as sketched in Fig.~\ref{fig:device} (a).
In a vertical layout, shown in Fig.~\ref{fig:device} (b), on the other hand,
molecules only physiosorb weakly on the graphene layer and the main
interactions are of van-der-Waals type.\cite{Hassan14, Silvestrelli14}
Extensive research has been conducted on the transmission features of
molecular junctions with bulk metal electrodes: Polarization and image charge
effects lead to a renormalization of the molecular levels and gateway states
arise due to hybridization with states located on the end group.
\cite{Neaton06, Peng09, Heimel13, Hüser15} The level alignment forces the
Fermi level of the electrodes, $E_F$ to be located in the gap between the
highest occupied and lowest unoccupied molecular orbital (HOMO-LUMO gap) and
transport is usually referred to as HOMO- or LUMO-dominated, depending on the
exact position of $E_F$.\cite{Neaton14} Whether or not these concepts can be
transferred to other junction layouts is not obvious and will be the focus of
our discussion.

The transport properties of molecular junctions with graphene as a top contact
layer in a vertical device setup have not yet been understood. The questions
that we would like to address in this paper are: How do electrons inject from
the metal electrode through the graphene into the molecule? What is the
signature of the molecule and how does it compare with the well-known case of
gold electrodes? And: how stable are results with respect to defects in
graphene and variations in the junction geometry?

Even though most experiments are using SAMs, we focus on the transport
properties of single molecules in this work. This means that no intermolecular
interactions are present, thus allowing for isolating chemical trends from
the molecule itself and studying the underlying physics on a basic level.

\section{Method\label{sec:method}}
The supercell is modelled with 6 by 6 gold atoms in each layer of the
electrodes with 7 layers in total. The lattice constant is
$\unit[4.176]{\text{\AA}}$. In order to maintain a feasible cell
size, the graphene is slightly strechted with a lattice constant of
$\unit[1.476]{\text{\AA}}$ (this corresponds to a strain of approximately
4 \%). In this way, $6 \times 8$ elementary unit cells of graphene match
with the (111) gold surface. Periodic boundary conditions are applied in
all directions. The geometry is shown in Fig.~\ref{fig:TE_vacuum} (a) for a
Au-graphene-Au junction without molecule. Unless otherwise stated, the distance
between the graphene layer and the gold surface is $\unit[3.3]{\text{\AA}}$, as
determined in a geometry optimization using the vdW-DF2 functional.\cite{Lee10}

Molecules are placed perpendicularly to the surface with a distance of
$\unit[2.41]{\text{\AA}}$ to the graphene layer. We note that variations in
the position and tilting angle with respect to graphene do not change our
results appreciably.

All transport calculations have been performed using a standard nonequilibrium
Green's Functions technique based on Density Functional Theory (NEGF-DFT).
\cite{Taylor01, Xue02, Brandbyge02, Thygesen03, Strange08, Strange11-1}
Within the Landauer-B\"{u}ttiker formalism for coherent transport,\cite{Meir92}
the electron conductance at $\unit[0]{K}$ is given by the value of the
transmission, $\tau(E, V)$, at the Fermi level of the electrodes in units of
the quantum of conductance, $G_0$:
\begin{equation}
G = G_0 \left.\tau(E, V=0)\right|_{E=E_F}.
\label{eq:G}
\end{equation}
For numerical stability, we evaluate conductance values by\cite{Pauly08}
\begin{equation}
G = G_0 \int\!dE \, \tau(E) \left(- \partial f(E,T)/\partial E\right),
\label{eq:Gnum}
\end{equation}
in steps of $\unit[0.01]{eV}$ in an energy range from $\unit[-3]{}$ to
$\unit[3]{eV}$ for $T=\unit[300]{K}$. $f(E,T)$ is the Fermi distribution.

Electronic structures and transmission have been obtained with the GPAW
code\cite{Enkovaara10} using the generalized gradient PBE exchange-correlation
functional,\cite{Perdew96}. We have verified that other functionals, in
particular vdW-DF2, do not change the transmission significantly. The choice of
the right functional is only an important issue for geometry optimizations.
Diffuse basis functions up to double-zeta polarization with a confinement-
energy shift of $\unit[0.01]{eV}$ have been employed.\cite{Strange11-1}
Special care is taken for the $k$-point sampling in the in-plane directions of
the graphene layer, since the transmission turned out to exhibit a strong
$k$-point dependence. A detailed discussion is given in the Supporting
Information. The results presented here have been obtained from averaging over
$(6 \times 6 \times 1)$ $k$ points.

\section{Results\label{sec:results}}
\subsection{Gold-graphene-gold junctions\label{subsec:Au-graphene-Au}}
\begin{figure}[t]
\begin{centering}
\includegraphics[width=\columnwidth,clip=]{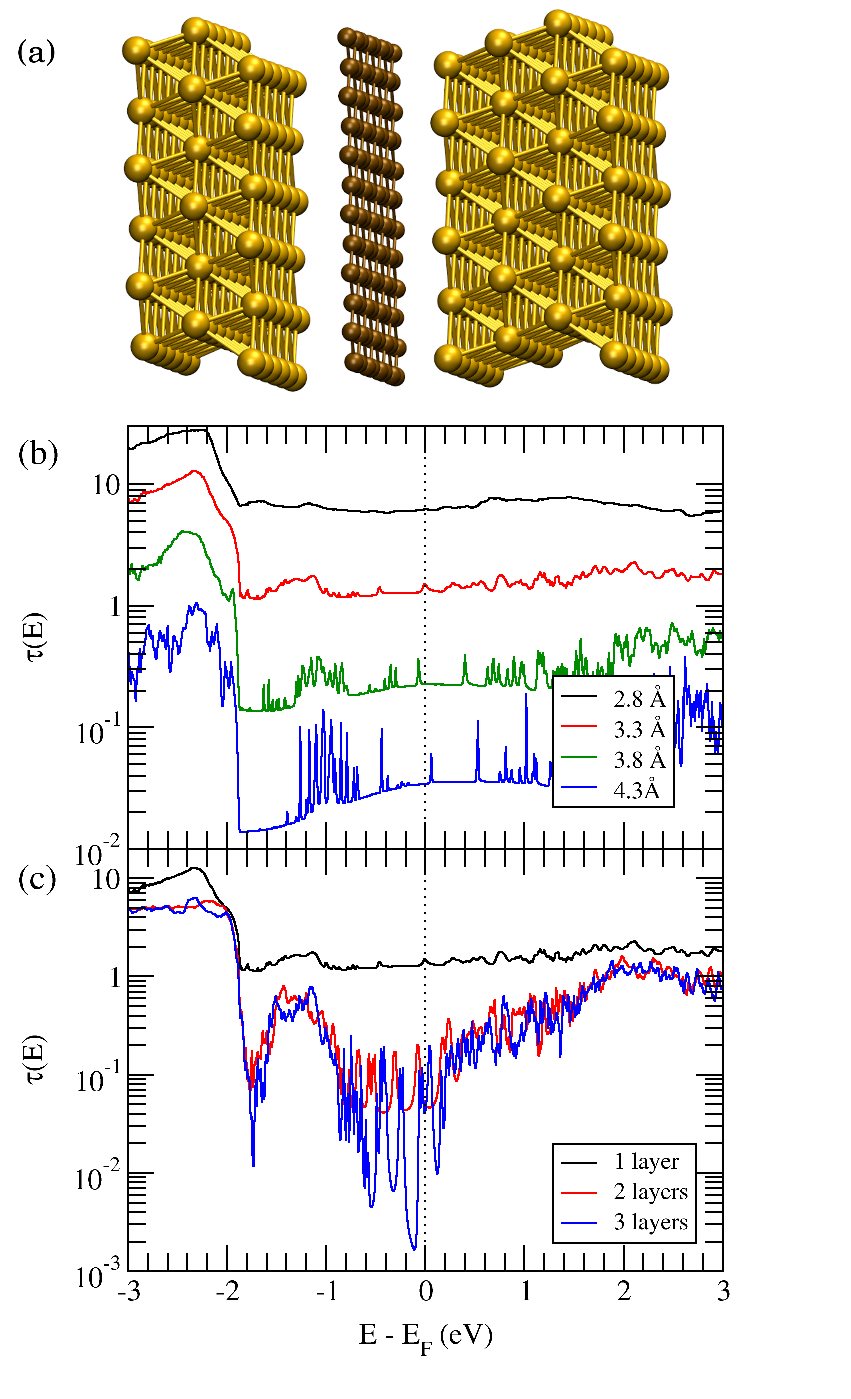}
\par\end{centering}
\caption{(a) Schematic junction geometry showing all atoms of the unit cell.
Vertical transmission for gold-graphene-gold junctions with
(b) varying distance and (c) different number of graphene layers.}
\label{fig:TE_vacuum}
\end{figure}

In order to get a first understanding of our new junction setup,
we calculate the vertical transport through a single graphene layer
symmetrically sandwiched between the gold electrodes for different distances
to the Au surface. The resulting transmission curves are shown in
Fig.~\ref{fig:TE_vacuum} (b). First, we notice that the signal is not entirely
smooth but exhibits a series of sharp resonances. This is a consequence of the
finite $k$-point sampling in our calculations.\cite{Thygesen05, Falkenberg15}
Nonetheless, they clearly
demonstrate the presence of graphene states, which are uncoupled or only very
weakly coupled to the gold leads. For an infinite number of $k$ points, the
resonances are expected to average out. The overall properties of the
transmission are a very high and broad peak around $\unit[-2.3]{eV}$ relative
to the Fermi level, a smaller split peak at $\unit[-1.3]{eV}$ and increasing
values up to $\unit[2]{eV}$ (see the Supporting Information for more details).
Around the Fermi level, the transmission is rather flat.
By increasing the distance to the gold surface, the conductance drops off by
several orders of magnitude, in agreement with a model for offresonant
tunnelling between the gold leads.

In Fig.~\ref{fig:TE_vacuum} (c), we compare the transmission for monolayer,
bilayer and trilayer graphene with an interlayer separation of
$\unit[3.35]{\text{\AA}}$ (as in bulk graphite) and $\unit[3.3]{\text{\AA}}$
distance to the gold surfaces on both sides. More peaks appear for
additional layers, due to the higher number of electronic states.
However, inbetween the resonances, the transmission decays exponentially with
increasing thickness, corresponding to an increase of the tunnelling distance.

\subsection{Introducing graphene defects\label{subsec:defects}}
\begin{figure}[t]
\begin{centering}
\includegraphics[width=\columnwidth,clip=]{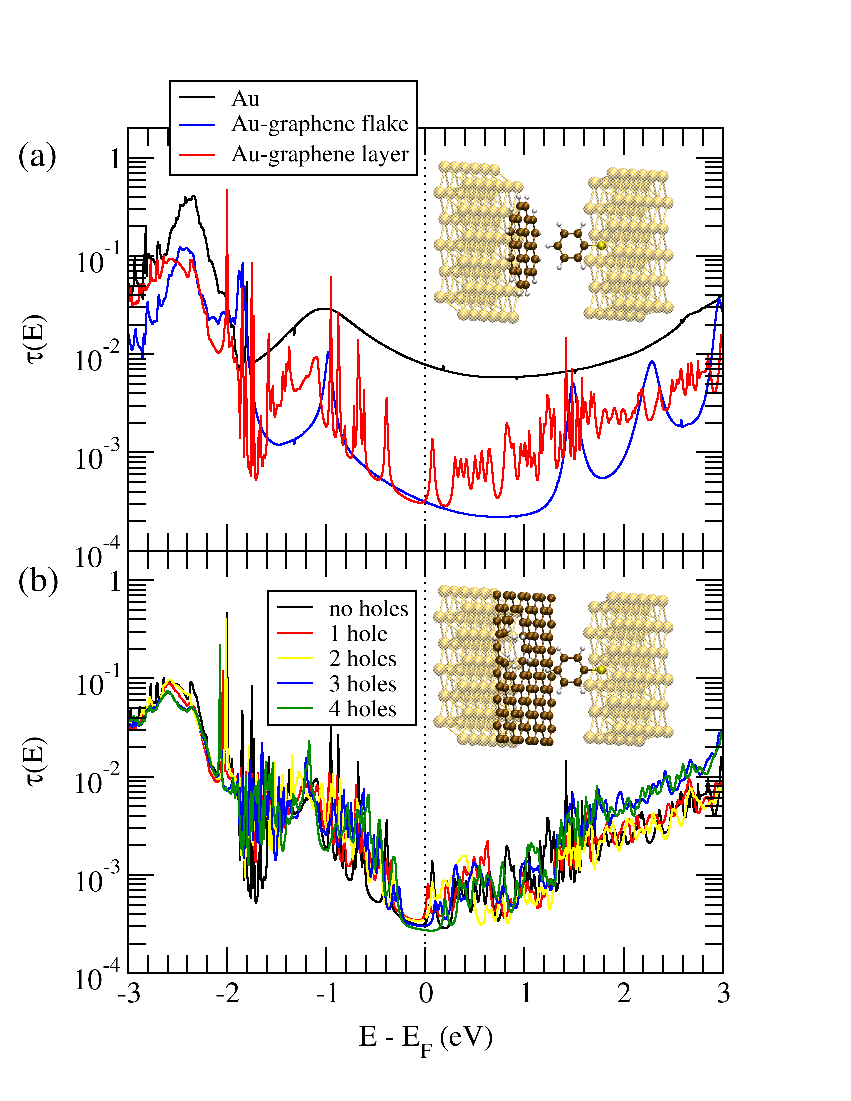}
\par\end{centering}
\caption{Transmission through (a) a benzene-thiol with gold leads only,
an additional graphene flake and an additional perfect graphene monolayer
and (b) a benzene-thiol with a graphene monolayer including hole defects.
Inset: Schematic junction geometry with (a) graphene flake and (b) 4 holes.}
\label{fig:benzene_defects}
\end{figure}
Using the example of a benzene-thiol molecule, we investigate the influence
of defects in the graphene layer. On the right side, the benzene is attached to
the gold electrode with a thiol end group binding to three Au atoms in a fcc
hollow site. Fig.~\ref{fig:benzene_defects} (a) shows the calculated
transmission curves for a junction consisting of the molecule with gold
electrodes only, an additional graphene flake (consisting of 7 rings and
passivated at the edges), and an additional perfect graphene monolayer. The
distance between the molecule and either the left gold electrode or the
graphene is $\unit[2.41]{\text{\AA}}$ in all cases. The black line (no
graphene) shows the typical sulfur-induced resonance around $\unit[1]{eV}$
below the Fermi level. Its height is reduced due to the asymmetric coupling.
Upon adding the graphene flake between the molecule and the left electrode, the
transmission drops by one order of magnitude. The peak at $\unit[-1]{eV}$
becomes sharper due to a weakened interaction with the gold electrode and
further resonances appear. The overall shape of the transmission, however, is
not altered significantly. This also holds when a perfect graphene monolayer is
inserted between the benzene-thiol and the left electrode. However, the signals
are obstructed by the features of the graphene itself. By simply looking at the
transmission, it is not clear what comes from the molecule and what from the
graphene. They cannot be separated.

As another type of defect, we introduce holes in the graphene layer by
removing 1 to 4 adjacent carbon atoms and passivating undercoordinated atoms
at the emerging edges. As plotted in Fig.~\ref{fig:benzene_defects} (b), the
resonance peaks move in energy (in accordance with a change in the density of
states of the graphene), whereas the value of the conductance and the curvature
in the vicinity of the Fermi level roughly remain stable.

Finally, we have varied the number of graphene layers. The observations are
similar to what is described for Fig.~\ref{fig:TE_vacuum} (c) in the previous
section for the case without molecule: A larger number of sharp resonances,
a reduction of the gap and a drastic lowering of the transmission in the gap.
The corresponding transmissions can be found in the Supporting Information.

\subsection{Varying the end group\label{subsec:endgroup}}
\begin{figure}[t]
\begin{centering}
\includegraphics[width=\columnwidth,clip=]{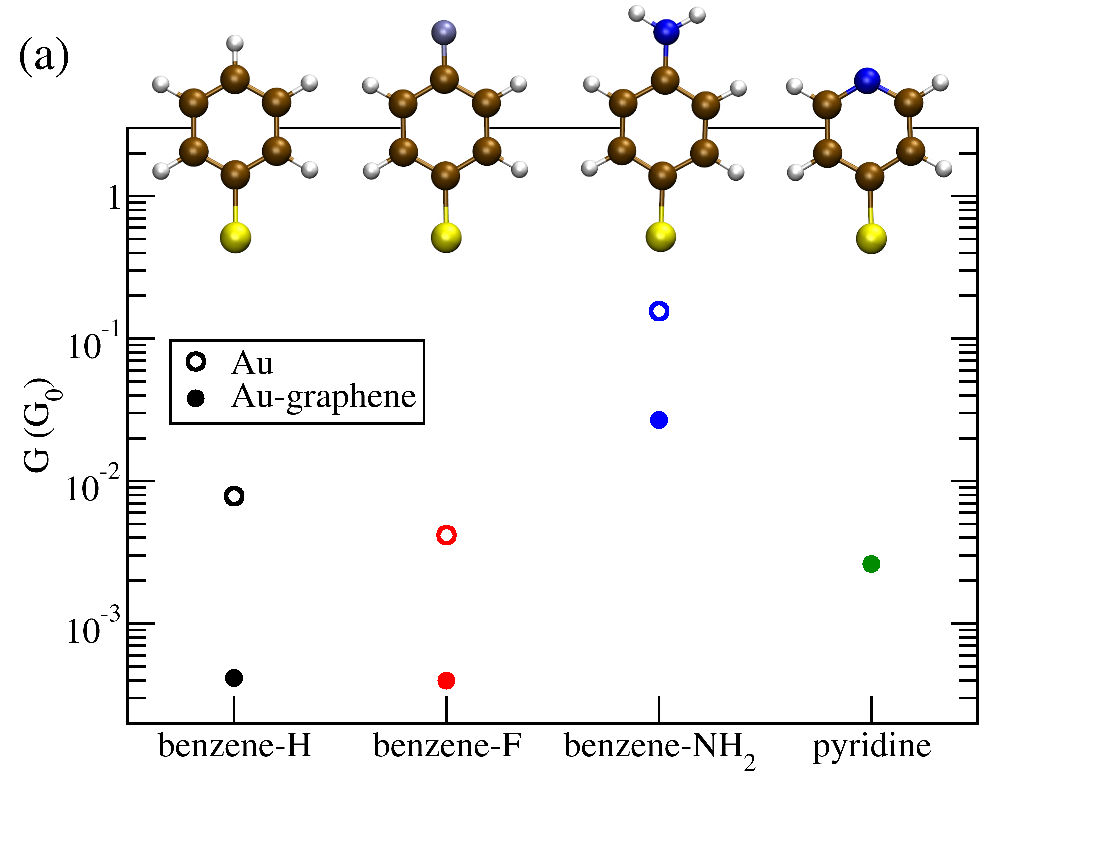}
\includegraphics[width=\columnwidth,clip=]{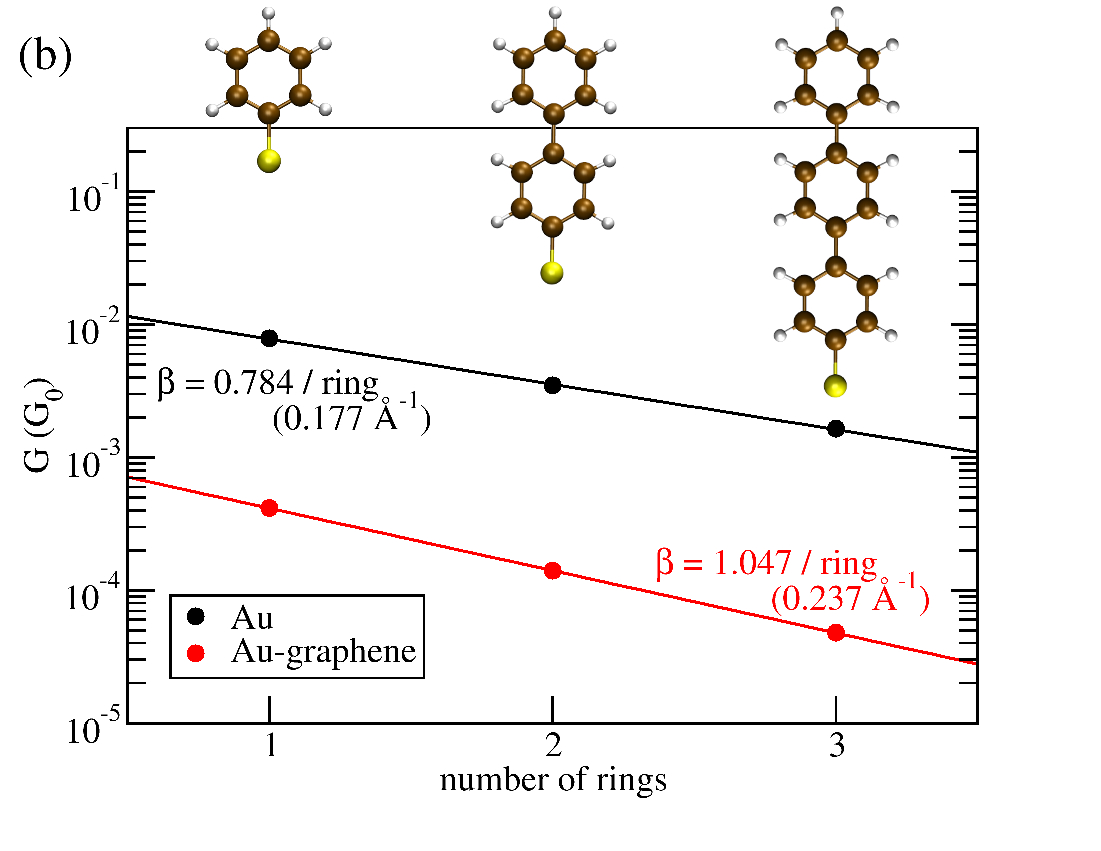}
\par\end{centering}
\caption{Conductance for (a) benzene-thiol with 3 different end groups
and pyridine-thiol and (b) $N$-phenylenethiols ($N = 1, 2, 3$)
and fitted values for the exponential decay factor, $\beta$.}
\label{fig:end_groups}
\end{figure}
For the gold-graphene-molecule-gold junction with a perfect graphene monolayer,
we explore the effect of different end groups on the transport properties and
thus the molecule-graphene interaction. The results are plotted in
Fig.~\ref{fig:end_groups} (a) for benzene-thiols with three different
terminating groups on the graphene side (hydrogen, fluorine and amine) and for
pyridine-thiol. First, we note that the corresponding transmissions (see
Supporting Information) for hydrogen and fluorine end groups are almost
identical. This indicates a very weak interaction. The two molecules are of the
same size and the conductance seems to be defined by the distance to the gold
electrode. For the amine end group, we see a broad resonance right below the
Fermi level. This arises from a conducting orbital formed by the nitrogen $p_z$
orbital and the $\pi$-system of the benzene ring. The conductance is much
higher than for the other molecules. It is important to point out that, also
here, the overall shape of the transmission curves is the same as for the case
of bare gold electrodes, but shifted to lower values and superposed by the
graphene signal.

Pyridine is a much shorter molecule and has a larger conductance than 
benzene-thiol with hydrogen and fluorine end groups. As for those two
molecules, there does not seem to be a conducting molecular orbital present in
the vicinity of the Fermi level. The level of ``background noise'' from
graphene states is the same in all cases. In fact, single resonances show up at
the same energies as for the junctions with a single graphene layer only (and
no molecule).

The relative differences in the conductance are about the same for the two
different junction setups. However, there are apparent differences in the
curvature of the transmission around $E_F$. This can be seen as a signature of
the molecule (including the end group).

For benzene-thiol with amine end group, we have also calculated the
transmission for increasing distances between the molecule and the graphene
layer. The only difference that we observe is a constant decrease, while its
shape remains unchanged (equivalent to Fig.~\ref{fig:TE_vacuum} (b)).
This is another indication for that there are no apparent interactions.

\subsection{Length dependence\label{subsec:length}}
A well known characteristic of a molecular wire is the exponential decay of the
conductance with length.\cite{Magoga97} In Fig.~\ref{fig:end_groups} (b), we
compare the length dependence for phenylenethiols with gold and gold-graphene
electrodes. The data points are fitted to an exponential function. The obtained
decay factors, $\beta$, are $\unit[0.18]{\text{\AA}^{-1}}$ and
$\unit[0.24]{\text{\AA}^{-1}}$, respectively. This is a small but nonetheless
apparent difference. However, we note that due to the ragged structure of the
transmission curves, the determination of the conductance values is sensitive
to numerical errors. Still, the qualitative picture remains.

\section{Conclusions\label{sec:conclusions}}
It is obvious that transport through vertical devices including graphene layers
differs substantially from that of gold-molecule-gold junctions. The appearance
of graphene signals in the whole energy range of the transmission makes it
difficult to identify signatures of the molecule itself. The classical picture
of HOMO- or LUMO-dominated transport does not seem to hold. Instead, one may
think of the graphene layer as behaving like a large molecule itself and the
signatures present in the transmission are those of both the molecule and the
graphene.

We note that the transmission has a very strong $k$-point dependence with
distinct, very sharp peaks for different $k$ points. Upon proper averaging,
their height is reduced and the transmission becomes smoother. For finite
$k$-point samplings, they cause a background noise signal. It is therefore more
meaningful to look at the conductance (evaluated for finite temperatures by
Eq.~(\ref{eq:Gnum})) and relative differences only.

For weakly physiosorbed molecules, no signs of interactions with the graphene
layer can be seen. The overall shape of the transmission, in particular the
curvature is not changed compared with junctions with gold only. This is best
seen in Fig.~\ref{fig:benzene_defects} (a) for going from no graphene to a
finite flake and finally to an infinite layer. Relative differences upon
changing the chemistry and length of the molecule are very similar.

In all cases, we see a drastic reduction of the conductance in agreement with
an increase of the tunnelling distance from gold to gold electrode. All our
observations hold when defects in the graphene are present. Measurements on
molecular junctions with a graphene top layer should therefore give very stable
results. Relative changes in conductance should be comparable to junction with
gold only.

For further understanding of the transport properties and exploration of the
transmission, measurements of $I-V$ curves and the thermopower might be very
helpful. Since the thermopower is proportional to the slope of the
transmission, it is much more sensitive to the presence and position of
sharp peaks close to the Fermi level. From our calculations, it was not
possible to extract numerically reliable values. Measured values, on the other
hand, could give valuable information on the nature of electron transport in
the molecular junctions and reveal trends across different classes of
molecules that are not available from the conductance only.
\cite{Karlström14, Hüser15} For the thermopower, larger and also qualitative
differences can be expected compared to regular gold-molecule-gold junctions.

\section{Acknowledgements}
This project has received fundings from the European Research Council
under the European Union's (EU) Seventh Framework Program (FP7/2007-2013)/ERC 
Grant Agreement No. 258806.

\section{Supporting Information}
The Supporting Information includes a detailed discussion of the $k$-point
sampling used in the calculation of averaged transmission curves as well as
transmissions for all molecules of this study.

\bibliography{bibfile}{}

\end{document}